\documentclass[amsmath,amssymb,floatfix,reprint,twocolumn,superscriptaddress,prb]{revtex4-2}

\usepackage{graphicx}
\usepackage{epsfig}
\usepackage{hyperref}
\usepackage{xcolor}
\usepackage{bm}

\begin{document}

\title{Power spectrum of magnetic relaxation in spin ice: anomalous diffusion in a Coulomb fluid}

\author{David~Billington}
\affiliation{School of Physics and Astronomy, Cardiff University, Queen's Building, The Parade, Cardiff, CF24 3AA, United Kingdom}
\author{Edward Riordan}
\affiliation{School of Physics and Astronomy, Cardiff University, Queen's Building, The Parade, Cardiff, CF24 3AA, United Kingdom}
\author{Clara~Cafolla-Ward}
\affiliation{School of Physics and Astronomy, Cardiff University, Queen's Building, The Parade, Cardiff, CF24 3AA, United Kingdom}
\author{Jordan Wilson}
\affiliation{School of Physics and Astronomy, Cardiff University, Queen's Building, The Parade, Cardiff, CF24 3AA, United Kingdom}

\author{Elsa Lhotel}
\affiliation{Institut N\'eel, CNRS $\&$ Universit\'e Grenoble Alpes, 38000 Grenoble, France}
\author{Carley Paulsen}
\affiliation{Institut N\'eel, CNRS $\&$ Universit\'e Grenoble Alpes, 38000 Grenoble, France}

\author{Dharmalingham Prabhakaran}
\affiliation{Department of Physics, Oxford University, Oxford, OX1 3PU, United Kingdom}
\author{Steven T. Bramwell}
\affiliation{
London Centre for Nanotechnology, University College London,
17-19 Gordon Street, London, WC1H 0AJ, United Kingdom}

\author{Felix Flicker}
\affiliation{
School of Physics, Tyndall Avenue, Bristol, BS8 1TL, United Kingdom}
\author{Sean~R.~Giblin}
\email{giblinsr@cardiff.ac.uk}
\affiliation{School of Physics and Astronomy, Cardiff University, Queen's Building, The Parade, Cardiff, CF24 3AA, United Kingdom}
\date{\today}

\begin{abstract}
Magnetization noise measurements on the spin ice Dy${}_2$Ti${}_2$O${}_7$ have revealed a remarkable `pink noise' power spectrum $S(f,T)$ below 4 K, including evidence of magnetic monopole excitations diffusing in a fractal landscape. However, at higher temperatures, the reported values of the anomalous exponent $b(T)$ describing the high frequency tail of $S(f,T)$ are not easy to reconcile with other results in the literature, which generally suggest significantly smaller deviations from the Brownian motion value of $b=2$, that become negligible above $T=20$ K. We accurately estimate $b(T)$ at temperatures between 2~K and 20~K, using a.c. susceptibility measurements that, crucially, stretch up to the relatively high frequency of $f = 10^6$ Hz. We show that previous noise measurements underestimate $b(T)$ and we suggest reasons for this. Our results establish deviations in $b(T)$ from $b=2$ up to about 20 K. However studies on different samples confirms that $b(T)$ is sample dependent: the details of this dependence agree in part, though not completely, with previous studies of the effect of crystal defects on monopole population and diffusion. Our results establish the form of $b(T)$ which characterises the subtle, and evolving, nature of monopole diffusion in the dense Coulomb fluid, a highly correlated state, where several dynamical processes combine. They do not rule out the importance of a fractal landscape picture emerging at lower temperatures where the monopole gas is dilute.
\end{abstract}

\maketitle

The spin ice material Dy${}_2$Ti${}_2$O${}_7$ has long been of principal interest among highly frustrated magnets. Large crystal field splittings combine with the large ($10\,\mu_B$) fields of the magnetic Dy${}^{3+}$ ions to create an excellent approximation to classical Ising spins. These occupy a pyrochlore lattice of corner-sharing tetrahedra, whose triangular faces lead to geometrical frustration in which the spins cannot simultaneously minimise their energies. Polarised neutron scattering at $T<1$ K reveals spin correlations characteristic of `ice rule' states in  which two spins point in, and two out, of each tetrahedron in agreement with a residual entropy measured by specific heat~\cite{bramwell:20}. Part of the excitement about spin ices is that the minimal excitations would take the form of mobile, localised sources and sinks of magnetization, resembling classical analogues of magnetic monopoles~\cite{castelnovo:08,Ryzhkin:2005aa}. Spin ice may be pictured as a dilute Coulomb gas of magnetic monopoles below $\sim 1$~K and an increasingly dense Coulomb fluid of both single and double charge magnetic monopoles above that temperature~\cite{Kaiser:13}.

A new tool to probe monopole dynamics came from theoretical proposals to use magnetic noise spectroscopy~\cite{Jaubert_2011,kirschner:18}, which passively probes the magnetic field fluctuations outside the sample. This technique allows a direct measurement of the noise spectral density $S(f,T)$. It has previously been applied to spin glasses, for example~\cite{Ocio:02}. The function $S(f,T)$ can alternatively be inferred from response measurements like a.c. susceptibility or neutron scattering, by means of the fluctuation dissipation theorem~\cite{kubo:66}. Methodology based on the response function is the standard approach, but it can be valuable to compare the two different measurements of $S(f,T)$ in cases of slow and complex dynamics, as afforded by spin glasses~\cite{Ocio:02} and spin ice~\cite{snyder:04,Matsuhira:12,Dasini24}, where the fluctuation-dissipation theorem may not apply~\cite{raban:22}. Here we focus on a temperature regime where we expect the fluctuation dissipation theorem to apply, so the two measures of $S(f,T)$ should be identical~\cite{morineau2024}.  

In the case of spin ice, numerical simulations suggested that magnetic monopoles would lead to a noise spectral density as function of frequency $S(f)\sim 1/f^b(T)$, with $1\le b \le 2$ varying
with temperature $T$, at frequencies above about $1\,$kHz~\cite{kirschner:18}. This `pink noise' would therefore be measurably distinct from the $b=2$ `red noise' expected for a simple paramagnet. This prediction was quickly confirmed experimentally using SQUID-magnetometry on Dy${}_2$Ti${}_2$O${}_7$, with pink noise measured at temperatures between $1-4\,$K, and frequencies up to $2.5\,$kHz. The anomalous $b(T)$ was found to decrease with increasing temperature, from $b=1.5$ at $T=1\,\textrm{K}$ to $b=1.2$ at $T=4\,\textrm{K}$~\cite{dusad:19}. A subsequent noise measurement in the temperature range $0.8-3.8\,$K, with frequencies up to $10^5\,$Hz, found close agreement, with $b=1.6$ at $T = 1\,\textrm{K}$ again decreasing to $b=1.2$ at $T = 3.8\,\textrm{K}$~\cite{samarakoon:22}. Recent theoretical work has proposed a mechanism by which $b=1.50$ can arise from the motion of isolated magnetic monopoles, relevant at the lowest temperatures of the spin ice regime (around $650\,$mK), as a result of their seeing an effectively fractal hopping environment~\cite{hallen:22}. 

However, these measurements have revealed a central mystery. Above $20\,$K it is well-established from neutron spin echo experiments that the classical spin ice materials based on Dy and Ho have $b=2$~\cite{Gardner_2011,Ehlers:2003aa}. While the data are more complete for Ho${}_2$Ti${}_2$O${}_7$~\cite{Ehlers:2003aa}, it is clear that Dy${}_2$Ti${}_2$O${}_7$ behaves in essentially the same way and on a similar temperature scale. Hence we can certainly expect a crossover to $b=2$ around $T=20$ K. Yet the noise-measured $b(T)$ are seen to \emph{decrease} from $b(1\,\textrm{K})=1.5$ with increasing temperature to a value $b(4\,\textrm{K})=1.2$, with little sign of the expected approach to $b=2$~\cite{dusad:19,samarakoon:22}. 

In this paper we confirm that, while showing some sample variation, $b$ is significantly closer to $2$ than implied by noise measurements and it evolves smoothly towards $b=2$ at higher temperature. A reason for systematic error in the noise-based estimates of $b$ could be a generic one: owing to the discrete time sampling and associated discrete Fourier transform, noise-based estimates of $S(f,T)$ measured in time are prone to distortions unless the sampling rate significantly exceeds the highest frequency component in the signal (an aspect of the Nyquist sampling theorem)~\cite{kirchner:05}. This condition cannot simply be achieved when $S(f,T)$ has a high frequency power law tail. The `aliasing' of the high frequency signal to lower frequencies then means that a log-log plot of the estimated $S(f,T)$ will show a point of inflection and a “flattened”, or pushed-up, high frequency tail. This leads to the possibility that the exponent $b$ could be under-estimated. This problem was  discussed by Kirchner~\cite{kirchner:05}, who proposed a method of spectral filtering to correct the noise based estimates, but that was not applied in the spin ice work. 

Therefore an aim of the present work was to accurately estimate the exponent $b(T)$. Linear response measurements like a.c. susceptibility and neutron scattering generally recommend themselves for this purpose, as, in contrast to noise based techniques, they should should yield unbiased estimates of $S(f,T)$ up to the highest measurable frequency, so long as the fluctuation-dissipation theorem applies and the susceptibility is linear (as is the case here). Although noise based methods face issues in estimating $b(T)$, they may still be useful, for example to test the fluctuation-dissipation theorem; or if noise itself is the subject of interest; or as a means of estimating $S(f,T)$ at low frequency.

A complication common to all measurement techniques is that $S(f,T)$ is sample-shape dependent in magnets. It can, however, be reduced to an intrinsic shape-independent function through the application of a demagnetizing correction (particularly important and subtle in spin ice)~\cite{Tweng:17}. In practice this is less easily applied to noise-based estimates of $S(f,T)$ than to susceptibility-based measurements, where the demagnetising correction can be applied directly to the complex susceptibility~\cite{Finger:77}.
\begin{figure}[t!]
\centerline{\includegraphics[width=1.0\linewidth]{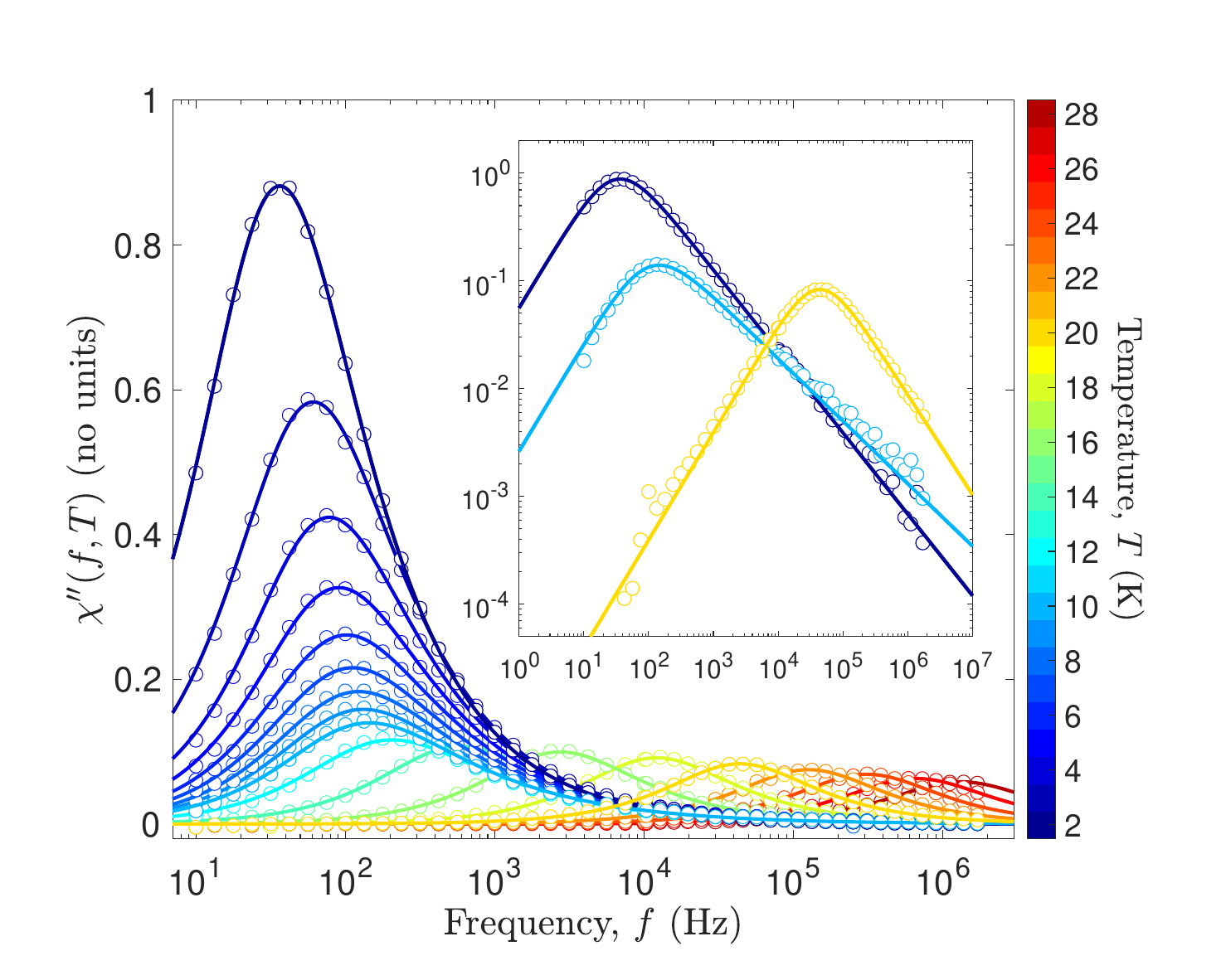}}
\caption{Imaginary part of the intrinsic, differential magnetic susceptibility, $\chi''(f,T)$, of single-crystal Dy$_{2}$Ti$_{2}$O$_{7}$ along the $[111]$ direction for various temperatures as a function of frequency. Inset: $\chi''(f,T)$ for selected temperatures ($T=2$, $10$ and $20$~K) on a $\log$-$\log$ scale to highlight the asymmetric deviation from Debye relaxation. The solid lines are fits to a Biltmo-Henelius function for $\chi''(f,T)$~\cite{Biltmo:2012aa}
}
\label{fig1}
\end{figure}
The a.c. susceptibility measurements were performed in a Quantum Design physical properties measurement system (PPMS). For measurements between $10^{1}\leq f\leq10^{4}$~Hz we employed the alternating current magnetic susceptibility (ACMS) option of the PPMS. For measurements in the range $f>10^{4}$~Hz we employed a unique, high-frequency magnetic susceptometer developed to operate within the PPMS cryostat up to frequencies $f\sim10^{6}$~Hz \cite{riordan:19}.  The extension to much higher than usual frequencies gives a crucial handle on the asymptotic power law behaviour of $S(f)$ at higher temperatures. In all cases data can be overlapped in the common frequency regime and demonstrated to be in the linear response regime for excitation fields between $0.04\,$mT -- $0.4\,$mT. The sample was zero-field cooled to the base temperature of the PPMS cryostat, $T=2$~K, and a.c. susceptibility measurements were performed with no static magnetic field as a function of increasing temperature for $2\leq T\leq28$~K. The Dy$_{2}$Ti$_{2}$O$_{7}$ single-crystal sample used in this study is the same as that studied previously in Ref.~\onlinecite{bovo:13}. For all measurements a demagnetization factor, $N=0.367$ was applied and the crystal was aligned along the $[111]$ direction with respect to the excitation field. The complex susceptibility was measured up to frequencies of 3 MHz, an order of magnitude higher than the highest frequency that could be resolved in the noise measurements at all measured temperatures ($2-22\,$K).

Starting with the complex intrinsic susceptibility $\chi'+i \chi''$, we obtained the function $S(f,T)$ by means of the classical fluctuation-dissipation theorem~\cite{kubo:66}: 
\begin{equation}\label{eq:fd}
S(f,T) = \frac{2k_BT}{hf}\chi''(f,T)
\end{equation}
where $h$ and $k_B$  are Planck's and Boltzmann's constants respectively. Fig.~\ref{fig1} shows the frequency dependence of $\chi''(f,T)$ measured for a distribution of temperatures for the Dy$_{2}$Ti$_{2}$O$_{7}$ sample. The inset shows a few temperatures for clarity on a log scale and the asymmetry demonstrates that, at the lower measured temperatures, the distribution is not well described by a single Debye relaxation process (which behaves as $\omega^1$ and $\omega^{-1}$ in the low- and high-frequency limits respectively). This phenomenon has been well studied in the context of spin glasses, where, as the spin-freezing temperature is approached from above, $\chi''(f,T)$ often exhibits appreciable asymmetry with respect to logarithmic frequency. Such behaviour is also observed in the dilute dipolar Ising magnet LiY$_{1-x}$Ho$_{x}$F$_{4}$ ($x=0.045$)~\cite{quilliam:08,biltmo:12}. 

\begin{figure}[t!]
\centerline{\includegraphics[width=1.0\linewidth]{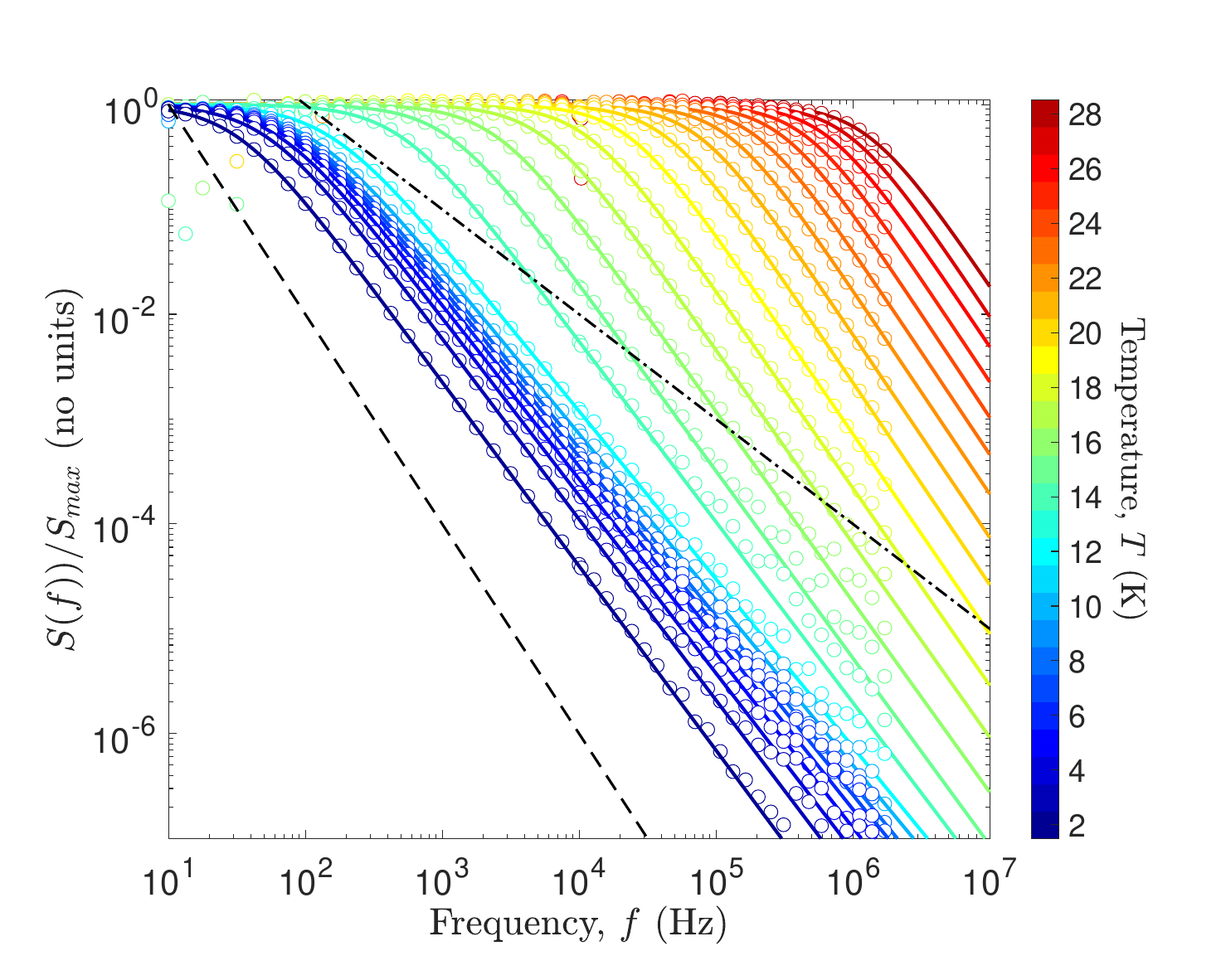}}
\caption{Normalized power spectrum of magnetic noise, $S(f,T)/S_{\rm max}(T)$ (from fitting Eq.~\eqref{eq:S}) as a function of frequency, for a single crystal of Dy$_{2}$Ti$_{2}$O$_{7}$ along the $[111]$ direction for various temperatures as a function of frequency on a $\log$-$\log$ scale. Solid lines are fits to Eq.\eqref{eq:fd}. The dash-dot and dash lines show $S(f) \propto f^{-1}$ and $\propto f^{-2}$ respectively. 
}
\label{fig2}
\end{figure}

The measured curves of $S(f,T)$ are shown in Fig.~\ref{fig2} where it is clear that the high frequency slope indicates an exponent $b(T)$ much closer to 2 than to 1 at all temperatures. We determined the exponent $b(T)$ roughly by fitting the high frequency gradient, and in more detail by fitting $\chi''(f,T)$ to a Biltmo-Henelius~\cite{Biltmo:2012aa} function and then transforming this to $S(f,T)$ to enable extraction of the exponent as described by:
\begin{equation}\label{eq:S}
S(f,T)=\frac{A\tau(T)}{1+(f\tau(T))^{b(T)}}.
\end{equation}
 This function can be derived from a restricted version of the the phenomenological Biltmo-Henelius (BH) $\chi''(f,T)$ originally introduced for spin glasses~\cite{biltmo:12}. It is seen to fit the data well. It is likely to give more accurate estimates of $b$ than the simple gradient method, particularly at high temperatures, where a limited frequency range above 100 kHz means that the estimate of $b$ will suffer from systematic errors. 

\begin{figure}[t!]
\centerline{\includegraphics[width=1.0\linewidth]{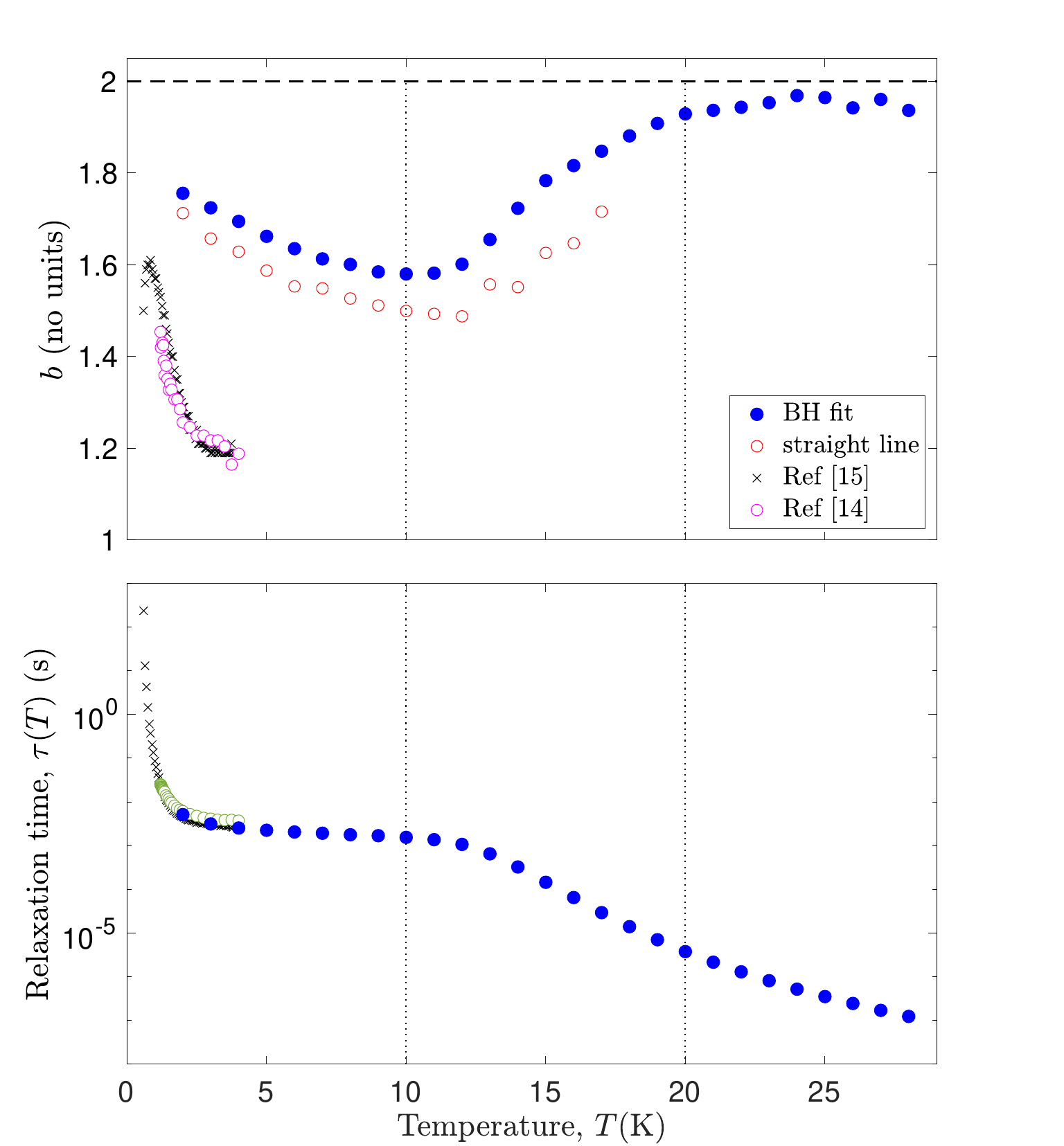}}
\caption{
({\it Top}) Temperature dependence of $b$ characterizing the distribution of relaxation times from the a.c. susceptibility with the exponents obtained from Eq.~\eqref{eq:S} and a straight line fit. Also plotted are the data obtained from the previous noise measurements. 
({\it Bottom}) Relaxation time as a function of temperature, $\tau(T)$, extracted from fits of $\chi''(f,T)$ to Eq.~\eqref{eq:S} compared with $\tau(T)$ form the published noise data on Dy${}_2$Ti${}_2$O${}_7$~\cite{dusad:19,samarakoon:22}. 
}
\label{fig3a}
\end{figure}

Results for the two estimates of $b(T)$ and the single estimate of $\tau$ are shown in Fig.~\ref{fig3a}, where they are compared with results estimated from the noise measurements of Refs.~\onlinecite{dusad:19,samarakoon:22}. It is clear that the noise measurements severely underestimate $b(T)$ while the susceptibility measurements connect the low temperature regime with the neutron scattering results~\cite{Gardner_2011}. As explained above, we expect estimates of $b$ based on the curve fitting method (Eq.~\eqref{eq:S}) to be more accurate than those based on the gradient method, but the two follow the same trend and their comparison at lower temperatures gives some idea of how the observed slope on the log-log plot differs from its asymptotic value.

In comparison to the estimates of $b(T)$ derived from the noise measurements~\cite{dusad:19,samarakoon:22}, the true $b(T)$ determined here remains anomalous but are always significantly closer to the Brownian motion value of $b = 2$ than the noise estimated value of $b = 1.2$, and agrees with the crossover to $b=2$ at $\approx$ 20 K observed by neutron scattering. The higher values of $b$ are consistent with the fact that monopole dynamics can be represented, on average, as Brownian motion in this regime. For example, based on measurements on the same crystal as used here, the mean mobility and mean diffusion constant (as defined with respect to the grand canonical ensemble) have been shown experimentally to accurately obey the Nernst-Einstein relation~\cite{bovo:13}. 

The thermal evolution of the relaxation rate derived from our fits suggests that effectively Arrhenius activated dynamics start to become relevant only above about 11 K. This agrees with the early susceptibility study of Matsuhira~\cite{Matsuhira:2001aa} and is consistent with neutron spin echo measurements, which found evidence for Arrhenius activation in Dy${}_2$Ti${}_2$O${}_7$ 
at temperatures of the order 20 K and above~\cite{Gardner_2011}. This is understood to be an Orbach-like phonon assisted spin flipping (or monopole hopping) associated with higher crystal field states~\cite{ruminy:2017}.

The absolute values of $b(T)$, however, show significant sample variation. Ref.~\onlinecite{Matsuhira:2001aa} reported $b\approx 1.5$ at 17 K, which contrasts with our value of $\sim 1.8$. In further work we measured $b(T)$ on several samples of varying quality and isotopic composition. Samples showed a similar temperature variation but with an absolute values of $b(T)$ dipping as low as 1.48, and varying by no more than $\sim$0.3 from our reported values at $T\sim 2$ K. The values of $b$ previously derived from noise measurements remain anomalously low, but they may be influenced by sample variation, in addition to the systematic error identified above. 

Sala \emph{et al}.~\cite{Sala:2014aa} showed how the precise relaxation rate and exponent $b$ can be affected by small concentrations of defects in the crystal structure, as these can affect the diffusion and density of monopoles. However, while they found that slower diffusion correlates with lower values of $b$, our measurements indicated the opposite, suggesting that there is still a lot to understand about the detailed role that defects play in monopole diffusion.  

Our results have implications for the measurement of anomalous noise exponents in magnetic systems, as well as for the physical processes in spin ice itself. 

For the estimation of the exponents $b(T)$, we have demonstrated the value of measuring susceptibility to high frequencies and we have identified systematic errors in the analysis of the noise experiments for $b(T)$. An obvious question is whether such errors also apply in the regime at $T< 1$ K where the anomalous value of $b$ is believed to be a signature of the effectively fractal hopping landscape for monopoles. In this regime the relaxation timescale slows considerably, and a number of processes are thermally frozen out. It might well be the case that the systematic errors on the exponent $b$ diminish accordingly. It would therefore be premature to rule out the validity of the noise measured $b$ in this regime -- there is a good chance that it remains correct, in accordance with the picture of monopole-hopping within a fractal landscape~\cite{hallen:22}. However, in general, for noise measurements, it would be appropriate to apply the filtering method of Kirchner~\cite{kirchner:05} to minimise, or perhaps even remove completely, certain systematic errors.

With regard to physical processes, it is important to clarify that the power spectral density, whether estimated by noise or susceptibility measurements, is not a spectroscopic probe of any individual dynamical process, but instead reflects the combination of all processes present: it can only distinguish these if they are well separated in frequency. In spin ice the magnetization relaxes by means of the thermal excitation and diffusion of both single and double charge monopoles~\cite{Jaubert:2009aa}. The actual dynamics involved include correlated quantum tunneling~\cite{tomasello:19} related to details of the local magnetic field, and phonon assisted spin flipping ~\cite{ruminy:2017}. In the dense monopole regime, monopole motion is highly correlated, partly as a result of local constraints~\cite{Jaubert:2009aa} and partly as a result of the many body Coulomb interaction~\cite{Kaiser:13}. A result of this complexity is that while, on average, monopole diffusion can be approximated as Brownian motion, in detail it differs from that. In the lower temperature regime ($T<1$ K) and at high temperature ($T>20$ K), the picture considerably simplifies as a more limited type of dynamics dominates in each case. While theoretical models~\cite{hallen:22,ruminy:2017,Paulsen:2016aa} have success in accounting for aspects of the dynamics in these limits, the evolution of the anomalous relaxation with temperature in the regime of 1 K - 20 K remains to be understood in detail. In accurately characterising the magnetic relaxation in this regime, we hope to facilitate future theories of this fascinating highly correlated state.

\begin{acknowledgments}
We acknowledge helpful conversations with M.~B.~Weissman and J.~N.~Hall\'{e}n. We gratefully acknowledge the financial support of the UK EPSRC, Grant Nos. EP/S016465/1 and EP/X012239/1. Research data from this paper will be made available via doi:10.17035/cardiff.27325305
\end{acknowledgments}

\bibliography{biblio_latest_24}

\end{document}